\documentclass[a4paper,fleqn]{cas-sc}

\usepackage[numbers]{natbib}
\usepackage{amsmath,amsfonts}
\usepackage{algorithm}
\usepackage{algcompatible}
\usepackage{amsfonts}
\usepackage{algpseudocode}
\usepackage{array}
\usepackage[caption=false,font=normalsize,labelfont=sf,textfont=sf]{subfig}
\usepackage{textcomp}
\usepackage{stfloats}
\usepackage{url}
\usepackage{verbatim}
\usepackage{graphicx}
\usepackage{makecell}
\usepackage{multirow} 
\usepackage{booktabs}
\usepackage[table]{xcolor}
\usepackage{svg}
\usepackage{tikz}
\usetikzlibrary{arrows.meta, positioning, shadows, calc}
\def\BibTeX{{\rm B\kern-.05em{\sc i\kern-.025em b}\kern-.08em
    T\kern-.1667em\lower.7ex\hbox{E}\kern-.125emX}}
\usepackage{balance}
\def\tsc#1{\csdef{#1}{\textsc{\lowercase{#1}}\xspace}}
\tsc{WGM}
\tsc{QE}

\begin{document}
\let\WriteBookmarks\relax
\def\floatpagepagefraction{1}
\def\textpagefraction{.001}

\shorttitle{LOIF: A New Sensitivity Factor}    

\shortauthors{D. Flores, Y. Sang, M. P. McGarry}  

\title [mode = title]{Line Outage Impact Factor (LOIF): A New Sensitivity Factor for Enhanced Transmission Observability}  



%

\author[1]{Daniel Flores}



\ead{dflores@umass.edu}



\affiliation[1]{organization={University of Massachusetts Amherst},
            addressline={151 Holdsworth Way}, 
            city={Amherst},
            state={MA},
            postcode={01003}, 
            country={U.S.A.}}

\author[1]{Yuanrui Sang}\cormark[1]


\ead{ysang@umass.edu}




\author[2]{Michael McGarry}

\ead{mpmcgarry@utep.edu}



\affiliation[2]{organization={The University of Texas at El Paso},
            addressline={500 W University Ave}, 
            city={El Paso},
            state={TX},
            postcode={79968}, 
            country={U.S.A.}}

\cortext[1]{Corresponding author}



\begin{abstract}
Transmission failures can lead to cascading failures and system blackout affecting millions of customers if not handled in time, and choosing the best locations to monitor the condition of the transmission system is crucial for power system reliability. In this paper, we propose a new sensitivity factor, the line outage impact factor (LOIF), which is especially useful for power system monitoring and can reveal the impacts of a transmission outage on the power flow of other lines more effectively than existing sensitivity factors, such as the line outage distribution factors (LODF). In this study, we apply the LOIF in transmission line outage detection in three test systems and compare it with LODF using a number of observed transmission line (OTL) selection methods based on these two sensitivity factors. Then we apply a machine learning algorithm to detect the outages of other lines by monitoring the selected OTLs, and the detection accuracy is evaluated using the F1-score. The results show that, in general, with the same number of OTLs, detection using the OTLs selected using LOIF achieved higher F1-scores. The pattern was especially consistent in large-scale systems, showing its potential in real-world applications.
\end{abstract}


\begin{highlights}
\item A new sensitivity factor, LOIF, is introduced to measure relative line outage impacts for monitoring.
\item Two methods for selecting monitoring locations to achieve maximum system coverage, Greedy MCP and high-$\eta$, are proposed and compared.
\item Results show that LOIF enables better monitoring location selection for line outage detection than LODF, validated up to a 1664-bus system.
\end{highlights}


\maketitle

\section{Introduction}
\label{sec:introduction}
\begin{figure}[pos=!htbp]
    \centering
    \includegraphics[width=0.5\columnwidth]{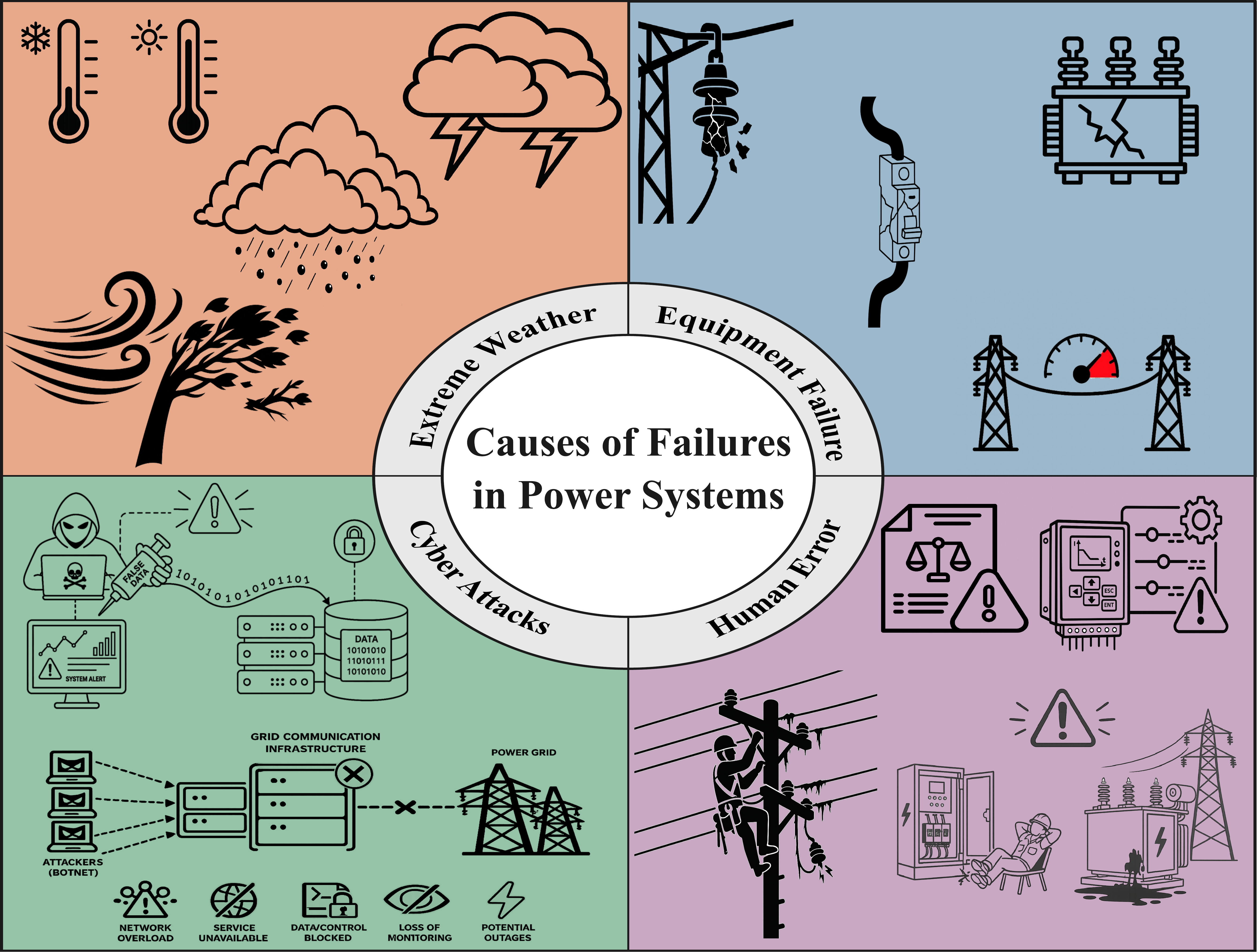}
    \caption{Causes of line outages categorized into four common groups: Extreme Weather, Equipment Failure, Cyber Attacks, Human Error}
    \label{fig:intro_fig}
\end{figure}

The power system is a complex engineered system that underpins the well-being of modern society, supporting industrial and commercial activity, healthcare services, and residential needs. 
As modern grids integrate emerging technologies, they also face more potential failure points. While a single component failure is often mitigated by protection schemes and operator responses, delayed or inappropriate actions may trigger cascading failures, as observed in historical blackouts \cite{zlotecka_characteristics_2018,venkatanagaraju_major_2024}. Power outages are commonly associated with weather and climate-related events, equipment failures, human error, and cyberattacks. Weather-related events, including extreme temperatures, thunderstorms, hail, and high winds, often cause physical damage to components such as transmission lines \cite{ren_analysis_2021}. Equipment-related outages can involve transmission line trips caused by overloading, short-circuit faults, insulation damage, relay failures, transformer disconnection, substation failures, and vegetation contact \cite{sharma_major_2021,vinogradov_analysis_2020}. Human-error-related outages may result from incorrect breaker switching, relay settings, maintenance, installation, regulatory compliance, or operational negligence \cite{ivanova_analysis_2023,ivanova_analysis_2025}. Cyberattacks can further compromise system operation through false data injection, denial-of-service attacks, or unauthorized control actions that may cause line and generator tripping \cite{tatipatri_comprehensive_2024,rajkumar_cyber_2023,rajkumar_dynamical_2024}. In addition, increasing penetration of renewable and inverter-based resources introduces uncertainty, variability, lower inertia, and new voltage, frequency, and protection challenges that may contribute to outages and cascading failures \cite{njoka_impact_2024,dudurych_impact_2021,impram_challenges_2020}. Figure \ref{fig:intro_fig} summarizes a number of factors that can cause failures in power systems.

Due to the complexity of system failures, it is essential to monitor power system operating conditions to prevent and mitigate the impacts of power outages. Power system monitoring is commonly performed using supervisory control and data acquisition (SCADA) systems and phasor measurement units (PMUs). PMUs can perform wide-area system monitoring by obtaining time-synchronized measurements of voltage and current phasors using a common time reference provided by GPS technologies \cite{penshanwar_phasor_2015}, and are becoming a more favorable tool than their counterparts, such as SCADA, as they can capture dynamic behavior at 30 to 120 samples per second, whereas SCADA can only observe steady-state behavior every 2-4 seconds \cite{biswal_real-time_2023}. PMUs have proven useful in multiple applications in power systems, such as state estimation, voltage and frequency stability, control and protection schemes, and the location and detection of system faults, such as line outages \cite{yu_comprehensive_2025,usman_applications_2019}. When a transmission outage occurs, neighboring lines and buses are often affected, and many line-outage detection methods have used PMUs to detect these changes by observing changes in voltage and current phasors, using pre-outage and post-outage data \cite{dai_line_2020,tate_line_2008,srikumar_ms_line_2015}. 

Traditionally, many line-outage detection methods were physics-driven, relying on common power system equations and mathematical approaches. These methods leverage PMU measurements, typically focusing on bus data such as voltage and current phasors. Simple methods such as \cite{alam_transmission_2021} simulate all possible outage scenarios, compute line and bus currents using AC power flow equations, and compare these simulated currents with collected PMU measurements to identify the outage with the lowest current mismatch. In \cite{jha_power_2022}, line outages are detected by analyzing the pre- and post-outage admittance matrices of the system, while considering limited PMU placement to reduce cost and measurement redundancy. Other approaches detect outages using residual-based methods such as weighted least squares (WLS) and weighted sum of squared residuals (WSSR) \cite{azizi_wide-area_2023}, where outages are identified by minimizing the mismatch between modeled disturbances and PMU measurements.

In recent years, line outage detection has been formulated as a classification problem using machine learning algorithms, aiming to classify outages by pattern recognition of system measurements. The idea is to use machine learning with effective features across the system (e.g., phase angle differences, power flow) that capture the system's behavior in a way that makes each outage condition discernible from others \cite{he_machine_2021,ibrahim_performance_2016}. In recent works, machine learning algorithms, such as least absolute shrinkage and selection operator (LASSO) and convolutional neural networks (CNNs), are used to detect outages, as the effects on the system are typically sparse and localized, meaning only a small subset of PMU measurements contains the most informative data \cite{yildiz_sparse_2025}. Other machine learning methods, such as k-nearest neighbors (kNN), have been used to detect outages, as in \cite{alam_kNN_2022}. Some machine learning methods use data such as system load forecasting and topology analysis to detect line outages, rather than relying solely on PMU measurements. In \cite{rogers_prediction-based_2024}, a method for line outage detection considers future loads by using Long Short Term Memory (LSTM) forecasting to capture future system behavior. Pairing this data with post-outage PMU measurements, machine learning methods such as support vector machine (SVM), Naive Bayes, and CNN are used for detection. In \cite{he_graph_2021} and \cite{chen_learning_2024}, graph-based neural networks are used to learn the system's topology, enabling the observation of the complex connectivity of system components for line outage detection. With PMUs, outages change measurements, and with topological features, the system's structure is also affected, providing additional information on different outage events.
The combination of physics-based and data-driven approaches using machine learning has also been explored, as in \cite{taghipourbazargani_machine_2023}, where modal analysis is used to extract features such as oscillation frequencies, damping ratios, and mode shapes. These features are then used to train machine learning classification algorithms, such as logistic regression (LR) and SVM, to classify outages.

Although PMUs are advantageous tools, they have high installation costs; hence, there is a need to optimally allocate PMU for full system observability while minimizing overall costs and ensuring effective data redundancy \cite{noureen_phasor_2017}. Existing methods for optimally allocating PMUs include leveraging zero-injection buses to observe adjacent buses \cite{rimal_optimal_2022} and using optimization techniques, such as mixed-integer linear programming-based models and heuristic methods \cite{more_literature_2013,johnson_critical_2021}. For example, genetic algorithms are used to determine which PMU locations to enhance the observability of a system \cite{yang_optimal_2020}. In addition, machine learning methods such as multinomial logistic regression (MLR) and LASSO can also be used for PMU placement \cite{kim_pmu_2018}. 
When monitoring transmission line outages, Line Outage Distribution Factors (LODF) can serve as a helpful tool for selecting the locations for monitoring devices, such as PMUs \cite {flores_transmission_2023}. However, we observed that LODF does not adequately capture the overall impact of a line outage \cite{flores_transmission_2025}, because the LODF only evaluates the percentage of power flow distributed to other lines when a line fails, and when a heavily utilized line goes out, its impact on other lines can be significant even if the LODF between the two lines is low. Thus, it cannot consistently distinguish between significant and insignificant outages, nor does it effectively differentiate outages that produce similar effects on the lines where the monitoring devices are located. This negatively affects the selection of sensitive lines that could be monitored to detect the outages of other lines.

\begin{figure*}[pos=!htbp]
\centering
\resizebox{\textwidth}{!}{
\begin{tikzpicture}[
    node distance=2.2cm,
    stage/.style={
        rounded corners=6pt,
        minimum width=2.8cm,
        minimum height=1.6cm,
        text=white,
        font=\sffamily\small\bfseries,
        align=center,
        drop shadow={shadow xshift=1pt, shadow yshift=-1pt, opacity=0.3}
    },
    arrow/.style={
        -{Stealth[length=8pt, width=6pt]},
        line width=1.5pt,
        color=gray!70
    },
    details/.style={
        font=\sffamily\scriptsize,
        align=left,
        text=black
    }
]

\node[stage, fill={rgb,255:red,41;green,98;blue,155}] (s1) {Power System\\Model};

\node[stage, fill={rgb,255:red,42;green,157;blue,143}, right=of s1] (s2) {Compute\\LOIF Matrix\\[2pt]{\normalfont\small$O_{a,b}$}};

\node[stage, fill={rgb,255:red,76;green,175;blue,80}, right=of s2] (s3) {OTL\\ Selection};

\node[stage, fill={rgb,255:red,230;green,126;blue,34}, right=of s3] (s4) {ML-Based\\Detection};

\draw[arrow] (s1) -- (s2);
\draw[arrow] (s2) -- (s3);
\draw[arrow] (s3) -- (s4);

\node[details, below=0.4cm of s1] (d1) {%
    \begin{tabular}{@{}l@{}}
    $\bullet$ Topology\\
    $\bullet$ Line flows\\
    $\bullet$ AC Power Flow Solutions
    \end{tabular}
};

\node[details, below=0.4cm of s2] (d2) {%
    \begin{tabular}{@{}l@{}}
    $\bullet$ Thresholding\\
    $\bullet$ Distinguishability\\
    \quad analysis
    \end{tabular}
};

\node[details, below=0.4cm of s3] (d3) {%
    \begin{tabular}{@{}l@{}}
    $\bullet$ Coverage sets\\
    $\bullet$ Greedy selection\\
    $\bullet$ Minimal OTL set
    \end{tabular}
};

\node[details, below=0.4cm of s4] (d4) {%
    \begin{tabular}{@{}l@{}}
    $\bullet$ Features selection\\
    $\bullet$ Outage classification
    \end{tabular}
};

\end{tikzpicture}
}
\caption{Overview of the proposed LOIF-based framework for transmission line outage detection.}
\label{fig:framework}
\end{figure*}
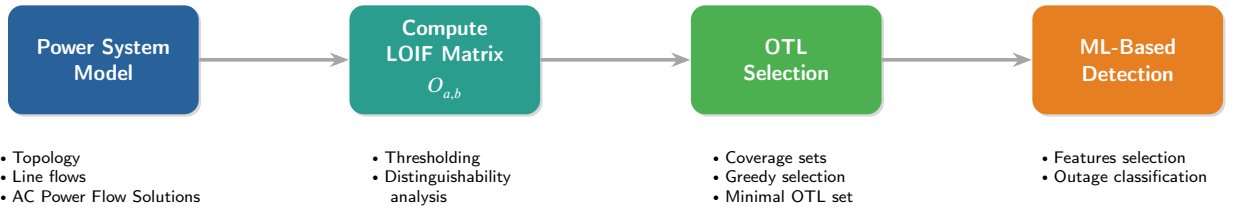

To address these limitations, this paper proposes an observability-driven framework for transmission line outage detection based on a new sensitivity factor and measurement selection strategy, as illustrated in Figure \ref{fig:framework}. The main contributions are as follows:

\begin{itemize}
    \item We introduce the \emph{Line Outage Impact Factor} (LOIF), a sensitivity factor that can effectively quantify the impact of a line outage on the power flow of another line from the perspective of the affected transmission line.

    \item We propose two LOIF-based methods, greedy maximum coverage (greedy MCP) and high-$\eta$, to identify monitoring locations, i.e., observed transmission lines (OTLs), that maximize system coverage and thereby enhance observability.

    \item We demonstrate, through experiments on IEEE 30-bus, 118-bus, and a large-scale 1664-bus system, that the proposed approach achieves comparable or improved line outage detection performance relative to LODF-based methods while requiring substantially fewer monitoring locations.
\end{itemize}

The rest of the paper is organized as follows. In Section \ref{sec:loif}, we define the line outage impact factor (LOIF) and describe its use to detect transmission line outages as a classification problem. Section \ref{sec:experimentalplan} presents the experimental plan to compare two OTL selection algorithms with random selection. The results of our experiments are analyzed and discussed in Section \ref{sec:results}. Finally, in Section \ref{sec:conclusion}, we summarize our conclusions and outline promising avenues for further investigation.

\section{Line Outage Impact Factor (LOIF)}
\label{sec:loif}



\subsection{From PTDF to LOIF: The Derivation Process}

This section presents the derivation of the Line Outage Impact Factors (LOIF), starting from Power Transfer Distribution Factors (PTDF), proceeding through LODF, and culminating in LOIF. This derivation provides a systematic sensitivity framework for analyzing the impact of transmission line outages.

\subsubsection{PTDF}

Under the DC power flow approximation, a linear relationship between nodal active power injections and transmission line flows can be obtained. The PTDF matrix, which is also called the injection shift factors (ISF) in some literature, is defined as a branch-by-bus sensitivity matrix. Each element of the PTDF matrix represents the sensitivity of the active power flow on a transmission line with respect to an incremental net active power injection at a bus, with all the injections absorbed by the slack bus \cite{wood1996power}.

Let \(H_{a,i}\) denote the PTDF of line \(a\) with respect to an incremental injection at bus \(i\). Then,

\begin{equation}
H_{a,i} = \frac{\Delta P_a}{P_i},
\end{equation}

where \(\Delta P_i\) is the change of active power flow on line \(a\) due to the net injection at Bus \(i\), and \(P_i\) is the net active power injection at bus \(i\). Therefore, the PTDF matrix has dimension \(n_l \times n_b\), where \(n_l\) is the number of transmission lines and \(n_b\) is the number of buses.

In matrix form, the PTDF matrix can be expressed as

\begin{equation}
\mathbf{H} = \mathbf{B}_f \mathbf{A} \mathbf{B}_{\theta}^{-1},
\end{equation}
where \(\mathbf{B}_f\) is a \(n_l \times n_l\) diagonal matrix, with the susceptance of each line listed on the diagonal of the matrix. \(\mathbf{A}\) is the adjacency matrix with a dimension of \(n_l \times n_b\) indicating the connection of the lines to each bus. The index of each row represents the line number, and the index of each column represents the bus number. In each row, an element has the value of \(1\) if the line is from the bus and \(-1\) if the line is connected to the bus. \(\mathbf{B}_{\theta}\) is the \(n_b \times n_b\) susceptance matrix under the DC power flow model, which neglects the conductance in the admittance matrix (\(\mathbf{Y_{bus}}\)) \cite{glover2023power}. 

\subsubsection{LODF}

The LODF quantifies how the flow on one line redistributes when the line is removed from service. Consider the outage of line \(b\), which connects buses \(i\) and \(j\). Let \(P_b^{\text{pre}}\) denote the pre-contingency active power flow on line \(b\). The LODF of line \(a\) with respect to the outage of line \(b\) is defined as \cite{guler2007generalized}

\begin{equation}
L_{a,b} = \frac{\Delta P_{a}}{P_b^{\text{pre}}},
\end{equation}
where \(\Delta P_{a}\) is the change in active power flow on line \(a\) caused by the outage of line \(b\).



\begin{figure}[pos=!htbp]
    \centering
    \includegraphics[width=0.5\columnwidth]{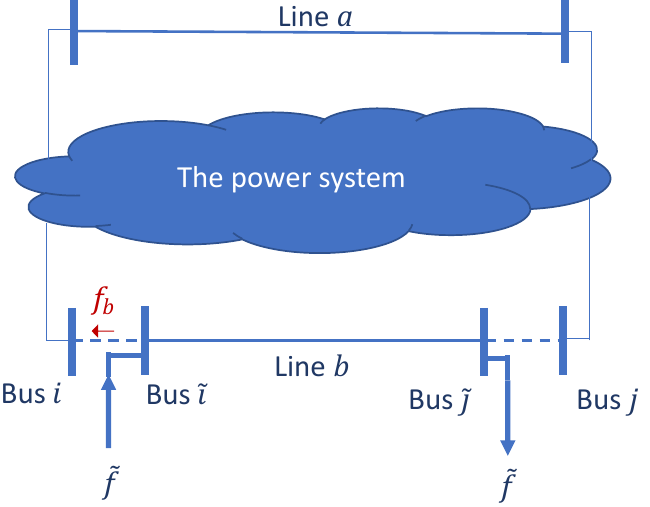}
    \caption{Simulating line outage with net power injections.}
    \label{fig:LODF}
\end{figure}

The derivation process for LODF starts with a simulation of a line outage using net power injections, as shown in Figure \ref{fig:LODF}. Assuming line \(b\) is the outaged line, and line \(a\) is the affected line, and we are interested in the LODF indicating how the outage of line \(b\) affects the power flow on line \(a\). To simulate the outage of line \(b\), we can add two buses, \(\tilde{i}\) and \(\tilde{j}\), which are very closely attached to buses \(i\) and \(j\), respectively. Then we inject a flow, \(\tilde{f}\), at bus \(\tilde{i}\), and withdraw the same amount of flow at bus \(\tilde{j}\), and the injection and withdrawal should have an impact of resulting a flow between buses \(i\) and \(\tilde{i}\) that has the same magnitude but opposite direction as the pre-contingency power flow on line \(b\), or \(f_b\). This will make the flow between buses \(i\) and \(\tilde{i}\) cancel out the original pre-contingency power flow. In this way, the outage of line \(b\) and its impact on other lines, such as line \(a\), can be simulated.

In this case, the LODF indicating how the outage of line \(b\) affects the power flow on line \(a\) can be calculated as

\begin{equation}
L_{a,b}
=
\frac{\tilde{f} H_{a,i} - \tilde{f} H_{a,j}}{\tilde{f} - \left( \tilde{f} H_{b,i} - \tilde{f} H_{b,j} \right)}
=
\frac{H_{a,i} - H_{a,j}}{1 - \left( H_{b,i} - H_{b,j} \right)}.
\end{equation}
in which the numerator represents the sensitivity of line \(a\) to a transaction from bus \(i\) to bus \(j\), while the denominator accounts for the self-effect of that transaction on the outaged line \(b\).



\subsubsection{LOIF}

While LODF captures the proportion of redistributed flow caused by a line outage, it does not directly reflect the relative significance of this change on the affected line. For example, the same MW change may be negligible on a heavily loaded line but significant on a lightly loaded line. To capture the relative impact of an outage on the affected line, the LOIF is defined as the normalized flow change on the affected line relative to its pre-contingency flow.

The LOIF of line \(a\) with respect to the outage of line \(b\) is defined as

\begin{equation}
\mathcal{O}_{a,b} = \frac{\Delta P_{a}}{P_a^{\text{pre}}},
\end{equation}
where \(P_a^{\text{pre}}\) is the pre-contingency active power flow on line \(a\). Using LODF, the LOIF can be calculated as

\begin{equation}
\mathcal{O}_{a,b}
=
\frac{L_{a,b} P_b^{\text{pre}}}{P_a^{\text{pre}}}.
\end{equation}

Using the PTDF-based expression for \(L_{a,b}\), LOIF can also be written as

\begin{equation}
\mathcal{O}_{a,b}
=
\frac{P_b^{\text{pre}}}{P_a^{\text{pre}}}
\cdot
\frac{H_{a,i} - H_{a,j}}{1 - \left( H_{b,i} - H_{b,j} \right)}.
\end{equation}

This final expression shows that LOIF combines two pieces of information: the network sensitivity captured by PTDF and LODF, and the relative flow magnitude between the outaged line and the affected line. Therefore, LOIF provides a normalized measure of how significantly the outage of line \(b\) affects the flow of line \(a\) compared to its own pre-contingency power flow.

\subsubsection{Post-Contingency Flow Estimation}

An advantage of LOIF is that it allows direct estimation of post-contingency line flows:
\begin{equation}
P_a^{\text{post}} = P_a^{\text{pre}} (1 + \mathcal{O}_{a,b}).
\end{equation}

\subsubsection{Discussion}

The derivation establishes a clear progression:
\begin{itemize}
    \item PTDF provides sensitivities of line flows to bus power injections.
    \item LODF quantifies flow redistribution after a line outage.
    \item LOIF incorporates pre-contingency operating conditions to measure the relative impact of outages.
\end{itemize}

Compared to LODF, LOIF captures both the magnitude of the outage flow and the pre-contingency operating condition of the affected lines. Since the affected lines are candidates to serve as OTLs, LOIFs are better suited for applications such as feature selection in machine learning-based transmission system monitoring and outage detection, where distinguishability and observability are critical.

\subsection{Using LOIF to determine the OTLs}
We define an OTL as a monitoring location, and a monitoring device, such as a PMU, is deployed either on the line or at a bus connected to the line to obtain power measurement data and provide features to solve the line outage detection problem. Ideally, we want to deploy such monitoring devices to monitor transmission lines whose power flows are sensitive to as many line outages as possible, so we can keep costs low by deploying as few devices as possible. 
LOIF values are useful to determine whether the flow of a line is sensitive to the outage of another line. For power measurement data from line $a$ to be used to detect a line outage on line $b$, two conditions must hold. First, the magnitude of the LOIF between OTL $a$ and a line $b$ experiencing an outage, $\mathcal{O}_{a,b}$, must exceed a threshold to ensure that the change of power flow on line \(a\) is discernible considering the impacts of noises and uncertainties in the measurements. Let $\beta$ be a threshold that represents the minimum magnitude that indicates a discernible power flow change. Condition one, $c_1(a,b)$, is therefore represented as: 

\begin{equation}
c_1(a,b) = \left| \mathcal{O}_{a,b} \right| \geq \beta \mbox{.}
\end{equation}

Second, the power flow change on line $a$ due to the outage of line $b$ must be discernible from the power flow change on line $a$ due to the outages of other lines. If we limit our consideration to only single line outages, then $\mathcal{O}_{a,b}$ needs only to be distinct among other LOIFs of line $a$ for outages of all the other lines. Let $\mathcal{L}$ be the set of all transmission lines, then for 
$\mathcal{O}_{a,b}$ to be distinct across all other possible line outages than that of line $b$, the magnitude of its difference from all the other LOIF values involving line $a$ must be greater than a threshold. Let $\gamma$ be that threshold, condition two, $c_2(a,b)$, is represented as:

\begin{equation}
c_2(a,b) = \forall_{c} \in \mathcal{L}\setminus \{a, b\} , \left| \mathcal{O}_{a,b} - \mathcal{O}_{a,c} \right|  \ge \gamma \mbox{.}
\end{equation}


Using these two conditions, we can derive the \textit{set of lines} whose outages could be detected by monitoring the power flow change of line $a$ as follows, and the set is also called the line observability set (LOS) in the rest of the paper.

\begin{equation}
S_{a}(\beta,\gamma) = \{ \forall_{b} \in \mathcal{L} \;\Big|\; c_1(a,b) \land c_2(a,b)\} \mbox{.}
\end{equation}

For each possible OTL $a$, we can use the size of set $s_{a}$ to define a coverage metric that measures how effective each OTL is for line outage detection. Let $\eta_a$ be the coverage metric for line $a$ and is defined to be the fraction of line outages that could be discerned by monitoring line $a$:
 
\begin{equation}
    \eta_a = \frac{|s_a|}{|\mathcal{L}|-1} \mbox{.}
\end{equation}

A simple algorithm for OTL selection can select the $x$ highest values of this coverage metric to determine a set of $x$ lines to deploy monitoring devices, such as PMUs, and we name this algorithm the \textbf{high-$\eta$} algorithm.


The parameters $\beta$ and $\gamma$ need to result in a low probability of mistaking one outage for any other outage or no outage. To formally
derive values that reduce that probability to an acceptably small value, we need to consider several factors:

\begin{enumerate}
  \item error in the estimates of LOIF due to the DC power flow simplification
  \item variability in power flow
  \item classification margin
\end{enumerate}

For this work, we use $\beta = \gamma = 0.1$ as a reasonable value to account for a small amount of estimation error and power flow variability.

\subsection{OTL selection as a maximum coverage problem}
\label{sec:mcp}
For optimal OTL selection, we want to select locations that maximize the coverage of the set of all line outages. The maximum coverage problem (MCP) 
takes several LOS, such as $S_a$, $S_b$, and $S_c$, etc., with potentially common elements, and a desired number ($x$) of them contain the maximum number of elements. Since each set indicates the coverage of that line, $x$ is also the number of OTLs selected. The MCP is known to be NP-hard; thus, a greedy algorithm is adopted to successively select sets that contain the largest number of uncovered elements. The very first set selected would be the one with the largest number of elements. Let $\mathcal{S}$ be the set of the $S_{a}$ sets for each line $a$, $\mathcal{C}$ be the set of currently selected sets, $\mathcal{R}$ be the set of currently covered elements, and $\mathcal{U}$ be the set of currently uncovered elements.
Algorithm \ref{alg:greedymcp} expresses this greedy algorithm to solve the MCP. We then select the lines associated with the $S_{a}$ subsets in $\mathcal{C}$ 
as the OTLs, we call this complete algorithm the \textbf{greedy MCP} algorithm.

\begin{algorithm}
\caption{greedy MCP$(\mathcal{L}, \mathcal{S}, x)$}
\label{alg:greedymcp}
\begin{algorithmic}[1]
    \State \textbf{Input:} Set of all lines: $\mathcal{L}$; collection of $s_{a}$ sets: $\mathcal{S}$; desired number of sets: $x$.
    \State \textbf{Output:} Selected sets: $\mathcal{C}$; coverage size: $|\mathcal{C}|$.

    \State $\mathcal{C} \leftarrow \emptyset$
    \State $\mathcal{R} \leftarrow \emptyset$
    \State $\mathcal{U} \leftarrow \mathcal{L}$

    \For{$i = 1$ to $x$}
        \If{$\mathcal{U} = \emptyset$}
            \State \textbf{break}
        \EndIf

        \State $S^* \leftarrow \arg\max_{S \in \mathcal{S} \setminus \mathcal{C}} |S \cap \mathcal{U}|$

        \State $\mathcal{C} \leftarrow \mathcal{C} \cup \{S^*\}$
        \State $\mathcal{R} \leftarrow \mathcal{R} \cup S^*$
        \State $\mathcal{U} \leftarrow \mathcal{U} \setminus S^*$
    \EndFor

    \State Return $(\mathcal{C}, |\mathcal{C}|)$
\end{algorithmic}
\end{algorithm}

\subsection{Classification model for line outage detection}
The \textit{features} of the classification model are the real and reactive power flows at the connecting buses (PF, QF, PT, and QT) of the $x$ 
selected OTLs (or PMU placements); $4x$ total features. The \textit{class label} is a single integer variable specifying the transmission line 
that is \textit{out-of-service}; with a value of 0 representing no transmission lines \textit{out-of-service}. Collected training data from a power 
system is used to train a machine learning classifier, such as the k-nearest neighbors (kNN) algorithm, to detect line outages from the power data 
collected by the monitoring devices at the $x$ selected OTLs.

\section{Experimental Plan}
\label{sec:experimentalplan}
\subsection{Objectives}
A set of experiments was conducted to compare the two OTL selection (or monitoring device allocation) algorithms with random selection. Labeled datasets were generated using \textsf{\textsc{Matpower}} simulations, and the kNN algorithm was used for classification. The labeled datasets were split into training and testing sets to evaluate the performance of the kNN classifier. The following OTL selection algorithms were compared using $F_1\text{-score}$ and statistical inference:

\begin{enumerate}
    \item \textbf{random} - randomly select $x$ OTLs,
    \item \textbf{high-$\eta$} - select $x$ OTLs with the highest coverage metric values, and
    \item \textbf{greedy MCP} - select $x$ OTLs using the greedy maximum coverage problem algorithm.
\end{enumerate}

\subsection{Labeled data set generation}
The labeled datasets were generated using stratified sampling across three different power systems with varying bus loads under $\pm5\%$ random load perturbations. 
The three different power systems have varying sizes:

\begin{itemize}
    \item IEEE 30-bus System
    \item IEEE 118-bus System
    \item Wisconsin 1664-Bus System
\end{itemize}

The widely-used \textsf{\textsc{Matpower}} electric power system simulator was used to simulate each of the three power systems with varying load 
conditions, as described above. The following procedure was used in each \textsf{\textsc{Matpower}} simulation to generate a class-balanced dataset for a specified power system and load condition: 
\begin{enumerate}
    \item Run AC Optimal Power Flow (OPF) to collect the generation of the system under normal conditions for base load and random load 
    variations (pre-outage generation).
    \item Under the same pre-outage conditions for load and generation, run AC Power Flow (PF) to observe the redistribution of power in the transmission network when disconnecting a single line. We run the PF solver rather than the OPF solver because we assume the system operator has not yet determined which line has gone out, so they do not redispatch the generators.
    \item When disconnecting a line, check for convergence when running the PF solution. 
    Real-world power systems are usually designed to meet the N-1 security requirement, meaning that the system should still be able to meet demand when any single line is removed. However, since real-world power system data is usually confidential, artificial test systems are used in this study, as in most power system research, and many of the systems are not N-1 secure. If the power flow problem does not converge when disconnecting a line, we assume that this system is not N-1 secure, which usually cannot happen in the real-world, and we neglect this line outage, skip collecting power flow data for that particular line disconnection, and focus on the line outages that do not lead to power imbalances in the system.
    \item Repeat steps 2 and 3 for all possible single-line outages.
\end{enumerate}


\subsection{Classification experiments}
Individual classification experiments were designed to vary the:
\begin{itemize}
    \item Feature Selection Tool - sensitivity factors used in the OTL selection algorithms: \{LOIF, LODF\}
    \item Algorithm - OTL selection algorithm, chosen from the set: \{ random, high-$\eta$, greedy MCP \}, and
    \item $x$ - number of OTLs, chosen from the set: \{ 1, 2, 4, 8, Full Coverage\} for small and medium sized systems and the set \{10, 20, 40, 80, Full Coverage\} for the larger 1664-bus system.
\end{itemize}
These experiments were repeated a total of 24 times, each with a different labeled data set, and conducted using the \texttt{scikit-learn} package for Python.

\section{Results and Analysis}
\label{sec:results}
From our experimental results, we compared the classification
performance between LOIF and LODF for different OTL
selection methods and with different numbers of OTLs. For
each system, we conducted 24 experiments, that is, 24 different
labeled datasets were used per system. In each experiment,
we ran a total of 30 tests (2 feature selection tools ×
3 OTL selection methods × 5 different OTL sizes) and collecting the average F1-score for each test. In Figs. \ref{fig:IEEE30_fixed}-\ref{fig:Wisconsin_sysem} of
this section, we compare the average F1-scores (of 24 total
experiments) when using LOIF and LODF as feature selection
tools. Each average F1-score is calculated with a confidence
interval of 95\%. We use the F1-score as our primary evaluation metric because it balances both precision and recall, providing a more complete measure of outage classification performance. In our experiments, the precision and recall trends closely matched the F1-score trends across all systems and OTL selection methods, so we focus primarily on the F1-score throughout this work.

\subsection{IEEE 30-Bus System}

\begin{figure}[pos=!htbp]
    \centering
    \includegraphics[width=0.6\columnwidth]{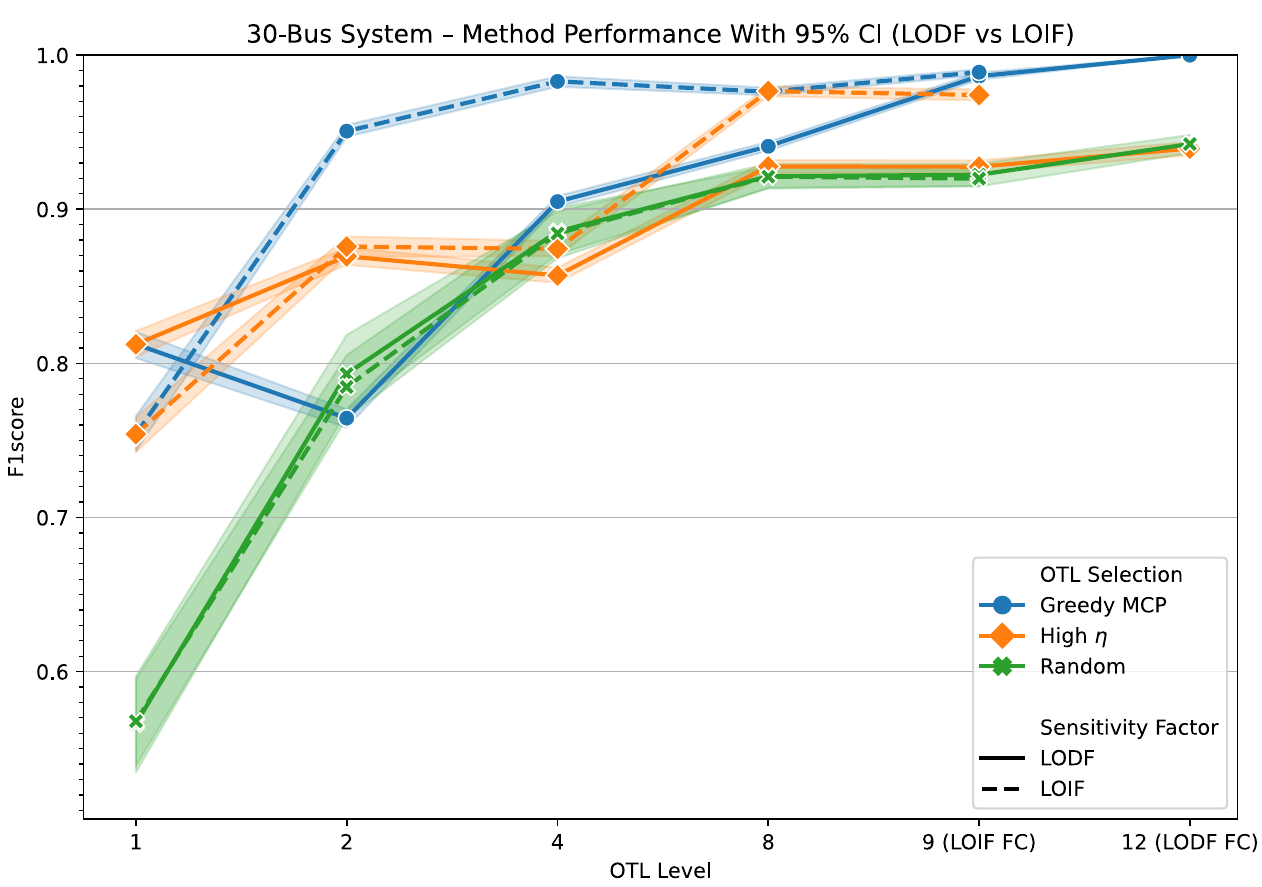}
    \caption{IEEE 30-bus system: Comparison in F1-score between LOIF and LODF across feature selection methods greedy MCP, high-$\eta$, and random for different numbers of OTL. $\beta$ and $\gamma$ are both set 0.1, and LOIF gets full coverage with only 9 OTLs while LODF gets full coverage with 12 OTLs}
    \label{fig:IEEE30_fixed}
\end{figure}

The IEEE 30-bus system contained a total of 41 transmission lines. Of these, 3 line-outage simulations did not converge in the power flow analysis, leaving 38 valid outage cases for consideration. When we had 1 OTL as shown in Figure \ref{fig:IEEE30_fixed}, LODF achieved better performance in the kNN classifier than LOIF, across both high-$\eta$ and greedy MCP methods. We obtained an F1-score of 0.81 with LODF for both methods and 0.75 with LOIF, giving us a 6\% difference in performance. When we increased the number of OTLs to 2, we saw that LOIF outperformed LODF. Specifically, for greedy MCP, we achieved an F1-score of 0.95 with LOIF, outperforming LODF by 19\%. For 4 OTLs, we observed that the kNN classifier achieved nearly perfect performance with an F1-score of 0.98 using the greedy MCP algorithm, with an 8\% improvement over LODF. There is a slight drop in performance when we use 8 OTLs in the greedy MCP method for LOIF, but it recovered once we reached full coverage at 9 OTLs, achieving a max average F1-score of 0.99. For LODF, it takes 12 OTLs to reach full coverage in the system, accounting for 31\% of all lines, and LOIF uses only 23\% of the system’s lines while still maintaining a relatively high F1-score. Overall, greedy MCP performed better than high-$\eta$ in selecting the best possible LOS to train the kNN model to classify all possible outages. As we increased the number of OTLs, the difference in performance between LOIF and LODF for greedy MCP was relatively low, due to the size of the 30-bus system. When an outage occurred, the change in power flow on the other lines is noticeable even with a small number of OTLs. 

\subsection{118-Bus System}

\begin{figure}[pos=!htbp]
    \centering
    \includegraphics[width=0.6\columnwidth]{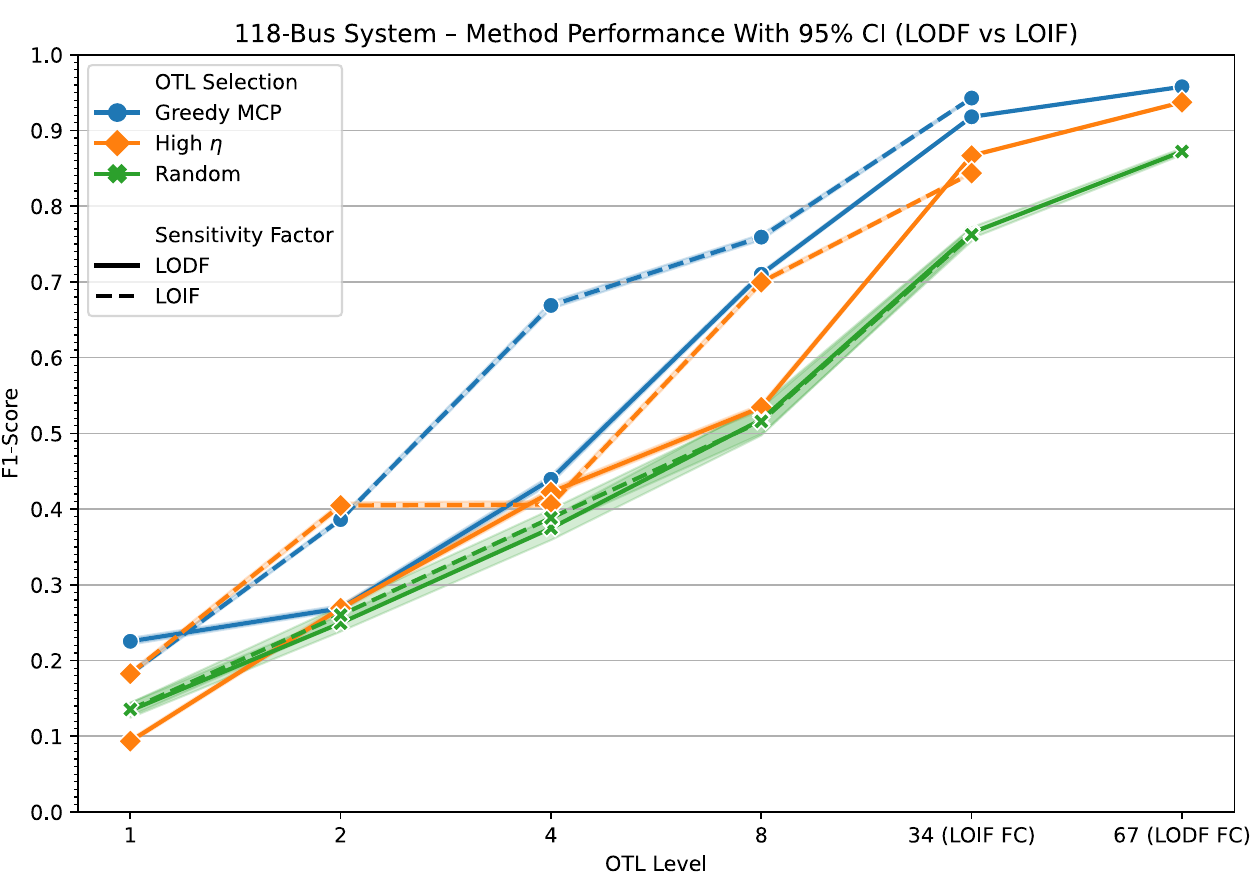}
    \caption{118-Bus system: Comparison in F1-score between LOIF and LODF across feature selection methods greedy MCP, high-$\eta$, and random for different numbers of OTL. $\beta$ and $\gamma$ are both set 0.1, and LOIF gets full coverage with only 34 OTLs while LODF gets full coverage with 67 OTLs}
    \label{fig:118bus_fixed}
\end{figure}

The IEEE 118-bus system contained a total of 186 transmission lines. Of these, 9 line-outage simulations did not converge, leaving 177 valid outage cases for subsequent analysis. For 1 OTL as shown in Figure \ref{fig:118bus_fixed}, the performance between LOIF and LODF varied across the three selection methods, with very low F1-scores ranging from 0.09 to 0.22. The performance in these cases is much poorer than those in the 30-bus system, because $1$ OTL is far from sufficient to make the entire system fully observable for a larger 118-bus system. For 2 OTLs, the F1-score nearly doubled for both high-$\eta$ and greedy MCP methods at 0.41 and 0.39, respectively, when using LOIF. LODF performs poorly compared to LOIF, achieving an average F1-score of only 0.27, resulting in a $10\%$ performance gap. With 4 OTLs, the benefits of using LOIF and greedy MCP to select effective features begin to become more prominent. The average F1-score increased to 0.66 for greedy MCP and LOIF, while the other combination of methods could only achieve a score of around 0.37 to 0.44. For 8 OTLs, LOIF and greedy MCP still produced the highest average F1-scores, jumping to a score of 0.76. As we attempted to achieve full coverage of the 118-bus system, we found that LOIF required only 34 OTLs, while LODF requires nearly twice of that, at 67 OTLs, and the difference in performance for greedy MCP is only about $2\%$. This means that LOIF offers greater distinguishability among all possible outages with almost $50\%$ fewer lines than LODF, resulting in a total usage of only $18\%$ of all transmission lines in the system. 

\subsection{1664-Bus System}

\begin{figure}[pos=!htbp]
    \centering
    \includegraphics[width=0.6\columnwidth]{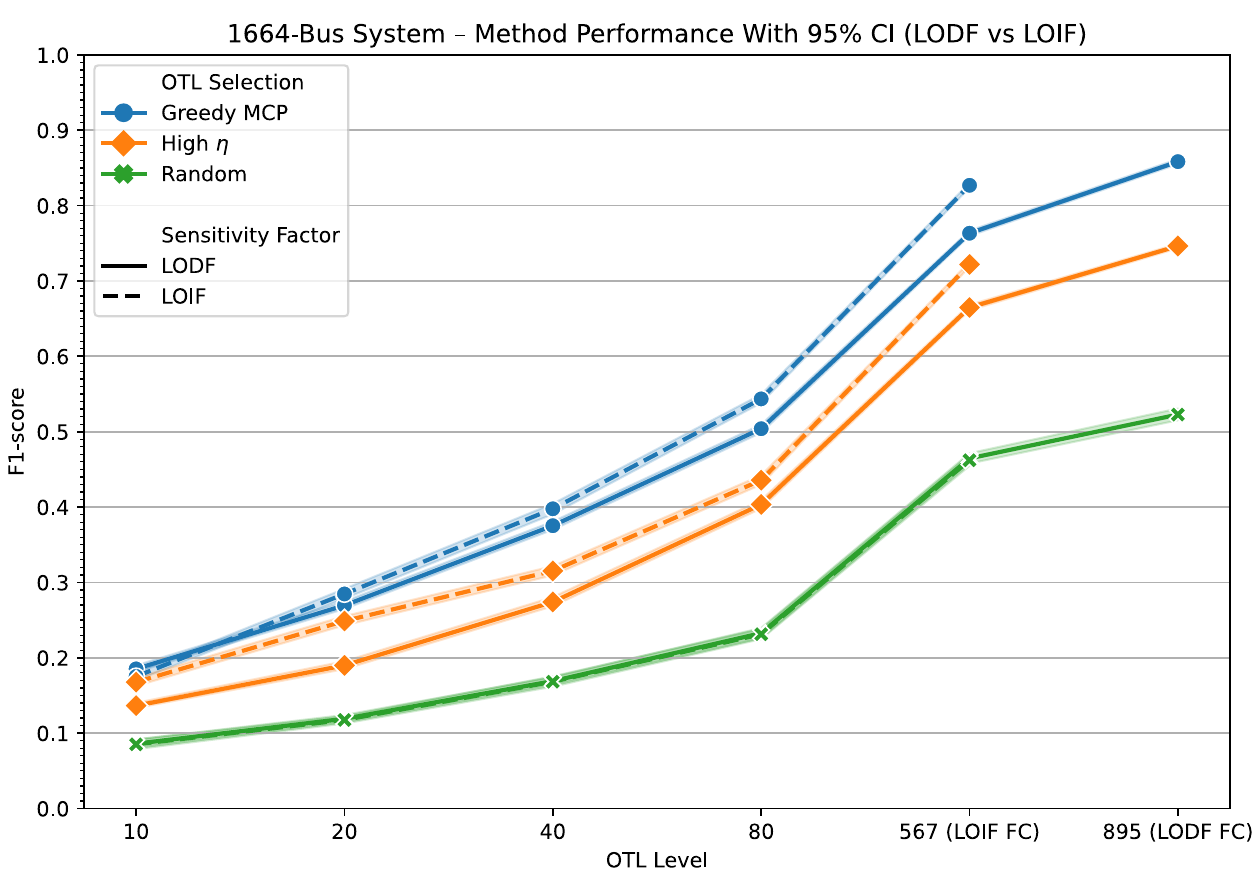}
    \caption{1664-Bus system: Comparison in F1-score between LOIF and LODF across feature selection methods greedy MCP, high-$\eta$, and random for different numbers of OTL. $\beta$ and $\gamma$ are both set 0.1, and LOIF gets full coverage with only 567 OTLs while LODF gets full coverage with 895 OTLs}
    \label{fig:Wisconsin_sysem}
\end{figure}

The Wisconsin 1664-bus system contained a total of 2,462 transmission lines. Of these, 374 line-outage simulations did not converge, leaving 2,088 valid outage cases for consideration. Since this test system is significantly larger than the former two, we used a larger number of OTLs to enhance its observability. As shown in Figure \ref{fig:Wisconsin_sysem}, the performance differences between selection methods, regardless of the sensitivity factors used, were clear, with greedy MCP outperforming the others. As we increased the number of OTLs, we observed a steady increase in the average F1-score across all three methods from 10 to 80 OTLs. From the results, it was also clear that LOIF performed better than LODF across all methods and OTL counts. To achieve full coverage of the system, LOIF required 567 OTLs compared to 895 OTLs for LODF, resulting in a 13\% difference in total line usage. For greedy MCP, we achieved a maximum average F1-score of 0.83 with LOIF and 0.86 with LODF for full coverage. Although LODF performed better, it required an additional 328 lines to achieve this, and the performance gap was only $3\%$.

Comparing the Wisconsin system to much smaller systems such as the 30-bus and 118-bus systems, we saw that the kNN performance decreased. This was because larger systems, such as the Wisconsin system, had greater line interconnectivity, resulting in a narrower spread of power flow redistribution during a line outage. In addition, most line outages redistributed power primarily within their local region, and the change in power flow decreased with OTLs farther from the outage. This resulted in features that were less distinguishable during certain outages, which, in turn, reduced the performance of the kNN classifier by introducing confusion when classifying all scenarios (normal conditions and all possible outages).

\subsection{The Advantage of LOIF}
With LOIF, we can formulate system observability as a greedy MCP problem by combining individual $S_a$ subsets to cover all possible outages across the three systems. This is because we can directly observe how the power flow at an OTL changes during different line outages, allowing us to construct subsets of outages that produce significant and distinguishable changes in power flow on each potential OTL. The way we derive each $S_a$ subset, along with greedy MCP, can also be applied using LODF, but outages with similar or small LODF values may still produce noticeable changes in power flow at the OTL. As a result, thresholding based on LODF may exclude outage conditions that still contain useful information for outage observability and detection. In Table \ref{tab:LOIF_adv}, we can see the potential of LOIF by observing the first 15 transmission line outages for each system at OTL $1$.

\begin{figure}[pos=!htbp]
    \centering
    \includegraphics[width=0.5\linewidth]{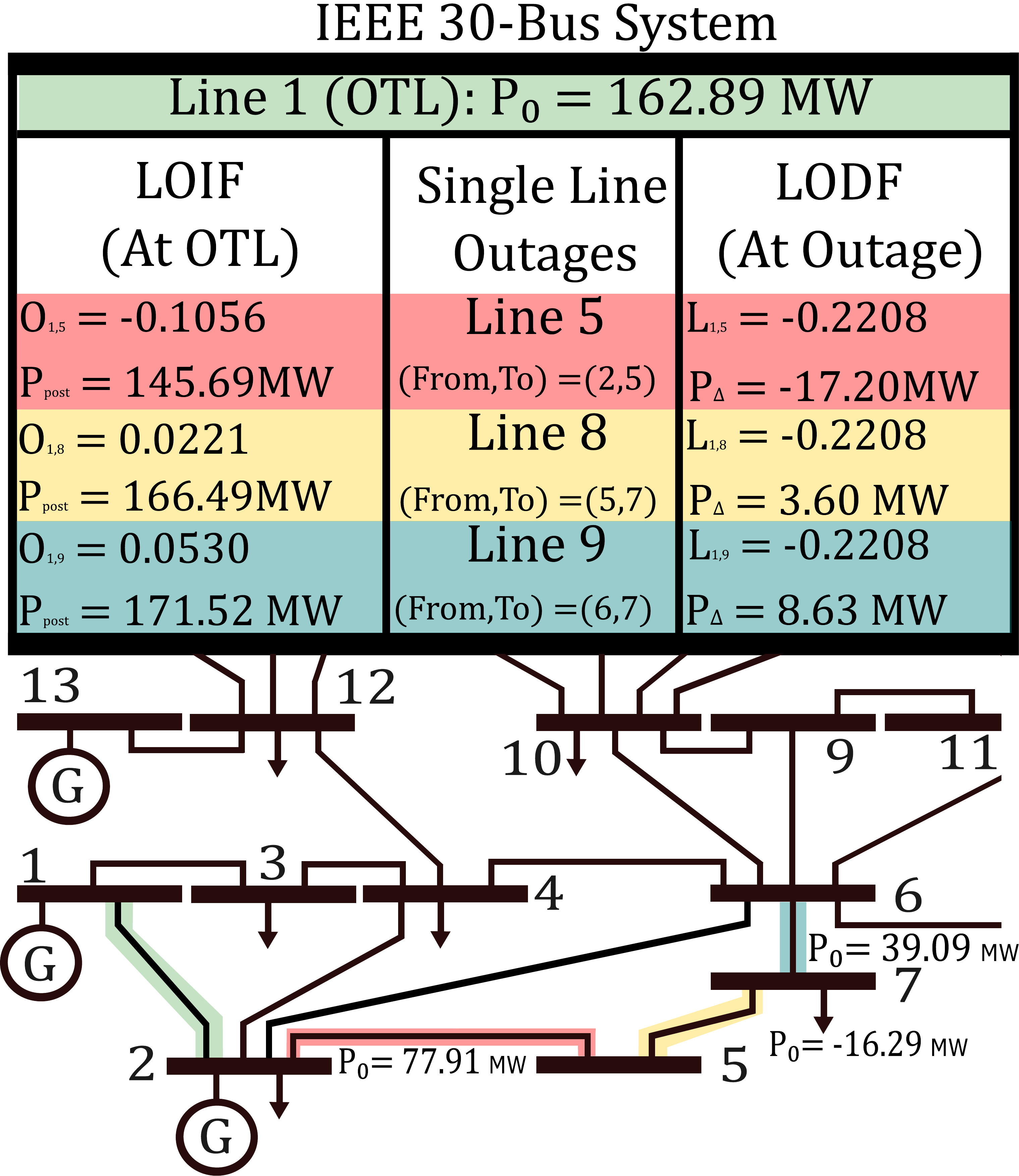}
    \caption{Example of similar LODF values producing different observable impacts at OTL 1 in the IEEE 30-bus system.}
    \label{fig:placeholder}
\end{figure}

This provides insight into why LOIF is preferred for system observability and why LODF may fall short in providing useful features. In the 30-bus system, lines 5, 8, and 9 all redistribute the same portion of their power flow, each with an LODF around $-0.22$. With LOIF, we can see that the changes across the three outages differ, resulting in relative changes of $-10\%$, $2\%$, and $5\%$ of the original power flow at OTL $1$. This phenomenon is further illustrated in Figure \ref{fig:placeholder}. A similar instance is shown in the 118-bus system across line outages 5, 6, and 15, each with an LODF of $0.0817$, but the power flow decreases by $60\%$, $24\%$, and $11\%$ of the original power flow at OTL $1$, respectively. Line outages 1 and 13 also share an LODF of $-1.0$, but outage 13 reduces the power flow at OTL $1$ by $268\%$ of its original value. In addition, there are instances in which a very small portion of the outage line’s power is redistributed, as in outages 4, 5, and 6 of the 1664-bus system, yet the change in power flow may still be noticeable. Outage 4 redistributes around $4\%$ of its power flow, but the change in power flow increases by $57\%$ of the original power flow at OTL $1$. The power flow increases by $14\%$ at OTL $1$ during outages 5 and 6, even though they redistribute only $2\%$ of their power. These outages across the three systems are only a small number of examples in which LOIF are better indicators of observability of line flows compared to LODF. It is possible to achieve good classification performance in smaller systems using LODF, given the small number of lines, but the performance quickly deteriorates as we increase the size of the system.

\begin{table*}[width=.9\linewidth,pos=!htbp]
\centering
\renewcommand{\arraystretch}{1.2}
\setlength{\tabcolsep}{4pt}
\caption{LODFs and LOIFs at OTL 1}
\begin{tabular}{
|c
|>{\columncolor{blue!10}}c >{\columncolor{blue!10}}c >{\columncolor{blue!10}}c
|>{\columncolor{purple!10}}c >{\columncolor{purple!10}}c >{\columncolor{purple!10}}c
|>{\columncolor{green!10}}c >{\columncolor{green!10}}c >{\columncolor{green!10}}c|}
\hline

\multirow{2}{*}{\shortstack{Outaged\\Line\\(b)}} 
& \multicolumn{9}{c|}{At OTL 1 (a)} \\ \cline{2-10}

& \multicolumn{3}{c|}{\cellcolor{blue!20}30-Bus} 
& \multicolumn{3}{c|}{\cellcolor{purple!20}118-Bus} 
& \multicolumn{3}{c|}{\cellcolor{green!20}1664-Bus} \\ \cline{2-10}

& $L_{a,b}$ & $\Delta P_{a,b}$ (MW) & $O_{a,b}$ 
& $L_{a,b}$ & $\Delta P_{a,b}$ (MW) & $O_{a,b}$ 
& $L_{a,b}$ & $\Delta P_{a,b}$ (MW) & $O_{a,b}$ \\ \hline

1  & -1.0000 & 162.89 & -1.0000 
   & \textbf{-1.0000} & \textbf{11.9159} & \textbf{-1.0000} 
   & -1.0000 & -44.2895 & -1.0000 \\ \hline

2  & 1.0000 & 82.75 & 0.5080 
   & 1.0000 & -39.0841 & 3.2800 
   & 0.0211 & 0.7281 & 0.0164 \\ \hline

3  & -0.3816 & -16.14 & -0.0991 
   & -0.0567 & 5.8422 & -0.4903 
   & * & * & * \\ \hline

4  & 1.0000 & 80.35 & 0.4933 
   & 0.4397 & -30.2121 & 2.5355 
   & \textbf{0.0444} & \textbf{25.2359} & \textbf{0.5698} \\ \hline

5  & \textbf{-0.2208} & \textbf{-17.20} & \textbf{-0.1056} 
   & \textbf{0.0817} & \textbf{7.0736} & \textbf{-0.5936} 
   & \textbf{-0.0192} & \textbf{6.0091} & \textbf{0.1357} \\ \hline

6  & -0.2939 & -17.27 & -0.1060 
   & \textbf{0.0817} & \textbf{2.8241} & \textbf{-0.2370} 
   & \textbf{-0.0192} & \textbf{6.0091} & \textbf{0.1357} \\ \hline

7  & 0.2924 & 21.22 & 0.1303 
   & * & * & * 
   & 0.0001 & 0.0074 & 0.0002 \\ \hline

8  & \textbf{-0.2208} & \textbf{3.60} & \textbf{0.0221} 
   & -0.0620 & -20.7589 & 1.7421 
   & -0.0005 & -0.0221 & -0.0005 \\ \hline

9  & \textbf{-0.2208} & \textbf{8.63} & \textbf{0.0530} 
   & * & * & * 
   & -0.0005 & -0.0276 & -0.0006 \\ \hline

10 & -0.0011 & -0.002 & -0.0002 
   & 0.0567 & 3.6291 & -0.3046 
   & * & * & * \\ \hline

11 & -0.0260 & -0.72 & -0.0044 
   & 0.0601 & 4.6024 & -0.3862 
   & 0.0001 & 0.0091 & 0.0002 \\ \hline

12 & -0.0201 & -0.33 & -0.0020 
   & 0.1129 & 4.0326 & -0.3384 
   & 0.0001 & 0.0091 & 0.0002 \\ \hline

13 & * & * & * 
   & \textbf{-1.0000} & \textbf{31.9159} & \textbf{-2.6784} 
   & 0.0001 & 0.0091 & 0.0002 \\ \hline

14 & -0.0260 & -0.72 & -0.0044 
   & 0.4069 & -3.8129 & 0.3200 
   & 0.0002 & 0.0129 & 0.0003 \\ \hline

15 & 0.0619 & -6.62 & 0.0161 
   & \textbf{0.0817} & \textbf{1.2714} & \textbf{-0.1067} 
   & * & * & * \\ \hline

\end{tabular}
\label{tab:LOIF_adv}
\end{table*}

\subsection{OTL Selection in the Greedy MCP and High-$\eta$ Methods}

In this section, we compare OTL differences between greedy MCP and high-$\eta$ across the three systems, using both LOIF and LODF. From Table \ref{tab:comparison}, we can see that using LOIF results in significantly fewer OTLs being selected for full coverage than using LODF. With greedy MCP and LOIF, we get full coverage with 9 OTLs for the 30-Bus system. When selecting the same number using high-$\eta$, there is a 55\% difference in the lines selected, but the difference in F1-score is only 0.01. With LODF, we get full coverage with 12 OTLs with a 58\% difference in lines selected and a difference of 0.06 in F1-score between greedy MCP and high-$\eta$. For the 118-bus and 1664-bus systems, we observe the same behavior, with OTL differences ranging from 47\% to 51\%, while F1-score differences are relatively low, ranging from 0.02 to 0.11. This shows that there are transmission lines that offer redundant information, allowing different OTL selections to produce similar classification performance.

\begin{table}[width=.55\linewidth,pos=!htbp]
\centering
\caption{Coverage Comparison of greedy MCP vs high-$\eta$}
\renewcommand{\arraystretch}{1.2}
\setlength{\tabcolsep}{3pt} 
\footnotesize 

\begin{tabular}{|l|cc|cc|cc|}
\hline
\multirow{2}{*}{System} 
& \multicolumn{2}{c|}{Coverage (\# of OTLs)} 
& \multicolumn{2}{c|}{Difference in OTLs (\%)} 
& \multicolumn{2}{c|}{$\Delta$ F1} \\
\cline{2-7}
& LOIF & LODF & LOIF & LODF & LOIF & LODF \\
\hline
30-Bus   & 9   & 12  & 55 & 58 & 0.01 & 0.06 \\
\hline
118-Bus  & 34  & 67  & 47 & 48 & 0.10 & 0.02 \\
\hline
1664-Bus & 567 & 895 & 49 & 51 & 0.11 & 0.11 \\
\hline
\end{tabular}

\renewcommand{\arraystretch}{1.0}
\label{tab:comparison}
\end{table}

The fact that the difference in F1-score between greedy MCP and high-$\eta$ is relatively small shows that
there may be more than one optimal set of OTLs, meaning there may be multiple solutions for full coverage, which could be explored in future work. 

\subsection{Analysis of kNN Confusion}

\begin{figure}[pos=!htbp]
    \centering
    \includegraphics[width=0.6\columnwidth]{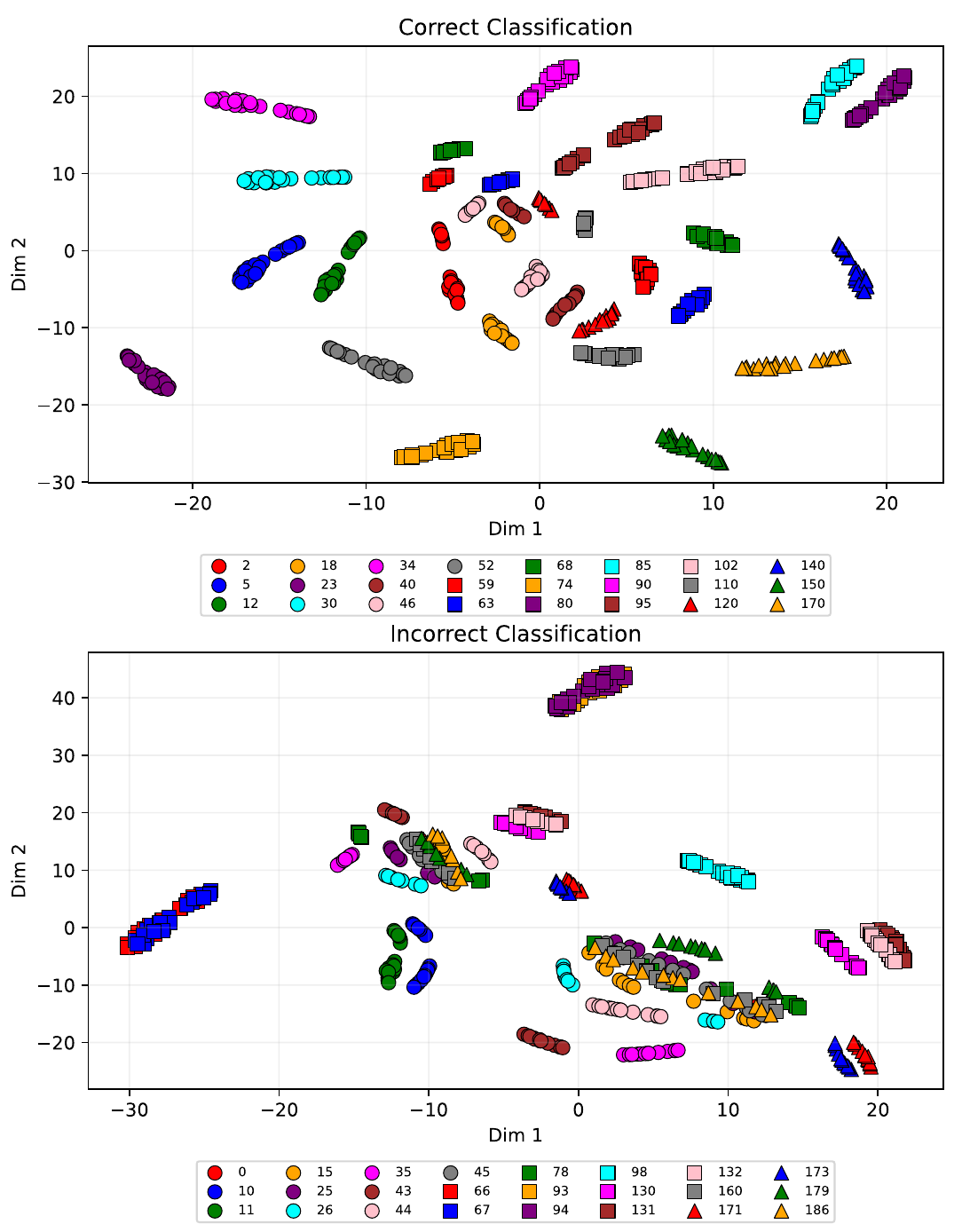}
    \caption{118-bus system: Comparison between power flow measurements for outage labels with correct kNN classification and outage labels with incorrect kNN classification.}
    \label{fig:test1}
\end{figure}

Because kNN is a distance-based classifier, 
overlaps in power flow measurements could confuse the classifier when determining the correct outage condition. To analyze this phenomenon, we observed the differences between correctly and incorrectly classified labels. We defined correctly classified labels as line outages with perfect classification scores (Precision, Recall, and F1-score = $1.0$), and incorrectly classified labels as outages with classification scores below $1.0$. We used t-distributed Stochastic Neighbor Embedding (t-SNE) to plot the features obtained from the 118-bus system in Figure \ref{fig:test1}, using OTLs selected via greedy MCP and LOIF for full coverage. Since we used active and reactive power flows at the connecting buses of our OTLs, we had a total of 136 features (4 × 34 OTLs), which were reduced into two dimensions in the figure. The top t-SNE plot shows outage labels with correct kNN classification, while the bottom plot shows outage labels with incorrect classification. From the plots, we can see that correctly classified outage labels formed more separated clusters, while incorrectly classified labels overlapped more in their power flow measurements. This overlap in power flow measurements made it difficult for the kNN classifier to distinguish between certain line outages. 

\section{Conclusion and Future Work}
\label{sec:conclusion}
In this paper, we present a detailed review of DC power flow-based power system sensitivity factors and introduce LOIF, which measures the impact of an outaged line on another line in the power system. In the case studies, we used LOIF as a feature-selection tool to detect transmission line outages using kNN in three test systems with different sizes and compared the detection F1-score across cases using two different sensitivity factors, i.e., LOIF and LODF, and three different OTL selection methods, i.e., greedy MCP, high-$\eta$, and random lines. The results show that LOIF required fewer OTLs than LODF to achieve full coverage while maintaining comparable detection performance.
In addition, greedy MCP generally produced higher F1-scores than high-$\eta$, with an average $50\%$ difference in the OTLs selected between the two methods. Although there was a large difference between OTLs, the performance gap was relatively small, suggesting that multiple sets of OTLs could be used to train the models while maintaining similar performances. By plotting the machine learning features using t-SNE, we showed that selected features that are more spread out in the high-dimensional space tend to yield good outage detection results.
In future work, we plan to test the detection methods with different machine learning classifiers, such as adaptive boosting decision trees, random forest, and extra trees, and develop refined methods for selecting the thresholds for \(\beta\) and \(\gamma\) to facilitate more effective OTL selection.










\printcredits

\bibliographystyle{cas-model2-names}

\bibliography{cas-refs}



\end{document}